# A proposal of implementing Hardy's thought-experiment with particles that never meet

Sofia Wechsler

Computer Engineering Center [1)]

**Abstract**

Hardy's thought-experiment is the strongest argument against the idea of continuous trajectories of particles from the source to the detector. This idea seems to result as an unavoidable conclusion from some simple experiments.
In its original form, Hardy's thought-experiment was proposed for two independent particles, a positron and an electron. The two either meet and annihilate, or **don't meet** but become entangled. An important feature of entanglements with particles that never meet is the difficulty they raise in front of the assumption of "hidden communication" between the particles. Indeed, if the particles don't meet, they can't have information of one another and it is difficult to explain how can they communicate.
In order to implement Hardy's paradox with particles that never meet, the present article proposes to use photons from three different lasers, one laser emitting UV rays, and the other two emitting at the wavelength of the signal photon, respectively the idler photon obtained from the down-conversion of the UV rays. The role of the annihilation in Hardy's experiment will be played by the destructive interference between a down-conversion pair and two photons from the low-energy lasers.

## 1. Introduction

The idea that particles follow continuous trajectories from the source to the detector not only is appealing to our classical intuition, but also seems to result as a conclusion from some single particle experiments. Indeed, a particle that impinges on a beam-splitter ends up either in the detector on the transmitted branch, $D_T$, or in the detector on the reflected branch, $D_R$, but never in both, as reported by A. Aspect and P. Grangier [2)]. A problem resulting from this experiment is as follows: let's remove the detector $D_R$, and assume that the detector $D_T$ clicks at some time $t_0$. It is certain that after this click the particle will not be on the reflected branch. However, reasoning from the point of view of an observer in movement with respect to the lab in the direction from $D_R$ to $D_T$, after the time $D_T$ clicked, the particle was not on the reflected branch beginning with a place closer to the beam-splitter than $D_R$. Considering observers in quicker and quicker movement, one eliminates places on the reflected branch closer and closer to the beam-splitter, and concludes that the particle followed a continuous trajectory from the beam-splitter to $D_T$ along the transmitted branch.

The terminology *empty/full waves* is encountered in the literature in conjunction with this situation. If the state of the particle at the outputs of the beam-splitter is described as $2^{-\frac{1}{2}}(|a_T\rangle + \iota|a_R\rangle)$, [3)] where $|a_T\rangle$ describes a wave packet traveling along the transmitted branch and $|a_R\rangle$ a wave packet traveling along the reflected branch, in this particular experiment $|a_T\rangle$ is said to be a *full wave* and $|a_R\rangle$ an *empty wave,* unable to impress the detector.

---

[1)] Nahariya, P.O.B. 2004, 22265, Israel
[2)] A. Aspect and P. Grangier, in Proc. Of NATO Advanced Study Institute on "Sixty-Two Years of Uncertainty", Aug. 5-15 1989 Sicily, Italy, edited by A.I. Miller Plenum Press, New York, page 45, (1990).
[3)] The Greek letter $\iota$ is used for sqrt(−1) in order to avoid confusion later in this text with the symbol i for the idler photon.

The idea of continuous trajectories and empty/full waves is not easy to reject. Models based on it can be found, that would explain the behavior of the polarization singlet contradicting the CHSH inequality. Such a model will be described in another article. It turns out that the strongest and simplest argument against the continuous trajectories is Hardy's proposed experiment. [4)]
Originally, this experiment was designed for two independent particles, a positron and an electron. They were supposed either to meet and annihilate or **not to meet and remain entangled**. However, due to the very low cross section of the annihilation, there exists a third possibility and it is dominant. The particles meet without reacting and remain non-entangled. Therefore the entangled pairs are extremely rare.

A recent simulation of Hardy's proposal was realized by Irvine et al. [5)] However they used pairs of down-conversion photons, so the particles meet at their very generation to the difference from Hardy's design. The significance of entangling never meeting particles was explained in the abstract. The present article suggests an implementation that fulfills this requirement. The next section describes the experiment, and section 3 discusses the contradiction regarding the trajectories' issue.

## 2. Hardy's experiment with never meeting particles

The proposed setup is shown in fig. 1. Two monochromatic lasers $L_S$ and $L_I$ emit beams of the same wavelengths as a signal, respectively idler photon obtained by down-converting the UV beam emitted by a third laser, $L_F$. The system of these three coherent beams is described by the product of three coherent waves, $|\alpha\rangle|\beta\rangle|\gamma\rangle$. The intensities of the beams from $L_S$ and $L_I$ are required to be equal s.t. $|\alpha| = |\beta|$. As it will appear in what follows they have to be very low, s.t. we may retain the first two terms in their development

(1) $|\psi\rangle = N (|0_S\rangle + \alpha|1_S\rangle)(|0_I\rangle + \beta|1_I\rangle)(\sum_{0}^{\infty} \gamma^n (n!)^{-\frac{1}{2}}|n_F\rangle)$ .

where $N$ is the normalization factor and the subscripts S, I, F, denote a photon from $L_S$, respectively from $L_I$, and from $L_F$. The beam-splitters BS split the beams from $L_S$ and $L_I$ according to

(2) $|1_m\rangle = 2^{-\frac{1}{2}}(|v_m\rangle + \iota|u_m\rangle)$ ,    $m = S, I$ .

The components $u_S$ and $u_I$ are redirected to illuminate the non-linear crystal $X$, on which also lands the beam from $L_F$ which is down-converted in the crystal. Two pinholes $P_S$ and $P_I$ in a screen behind the crystal select down-conversion photons of wavelengths equal to those from $L_S$ and $L_I$. Introducing (2) in (1) and considering the down-conversion process as described in the Appendix, one gets the pair-production term

(3) $(|v_S\rangle + \iota|u_S\rangle)(|v_I\rangle + \iota|u_I\rangle)|0_F\rangle + (2q\gamma/\alpha\beta)|0_S\rangle|0_I\rangle|u_S\rangle|u_I\rangle$

Let's put it in a more convenient form

(4) $(|v_S\rangle|v_I\rangle + \iota|v_S\rangle|u_I\rangle + \iota|u_S\rangle|v_I\rangle)|0_F\rangle - |u_S\rangle|u_I\rangle|0_F\rangle + (2q\gamma/\alpha\beta)u_S\rangle|u_I\rangle|0_S\rangle|0_I\rangle$.

By similarity with an experiment performed by Herzog et al.[6)] we assume that the last two expressions in (4) will interfere. The following additional constraint is required to be satisfied by

---

[4)] L. Hardy, "Quantum Mechanics, Local Realistic Theories, and Lorenz-Invariant Realistic Theories" Phys. Rev. Lett., Vol. **68**, no. 20, page 2981, (18 May 1992).
[5)] Irvine et al., "Realisation of Hardy's Thought Experiment", quant-ph/0410160v1, (2004).
[6)] T. J. Herzog et al., "Frustrated Two-Photon Creation via Interference", Phys. Rev. Lett., vol. **72**, no. 5, (31 Jan. 1994). For an analysis of the phenomenon see also S. Wechsler, "What was in the apparatus before the click?", quant-ph/0411039, (Nov. 2004).

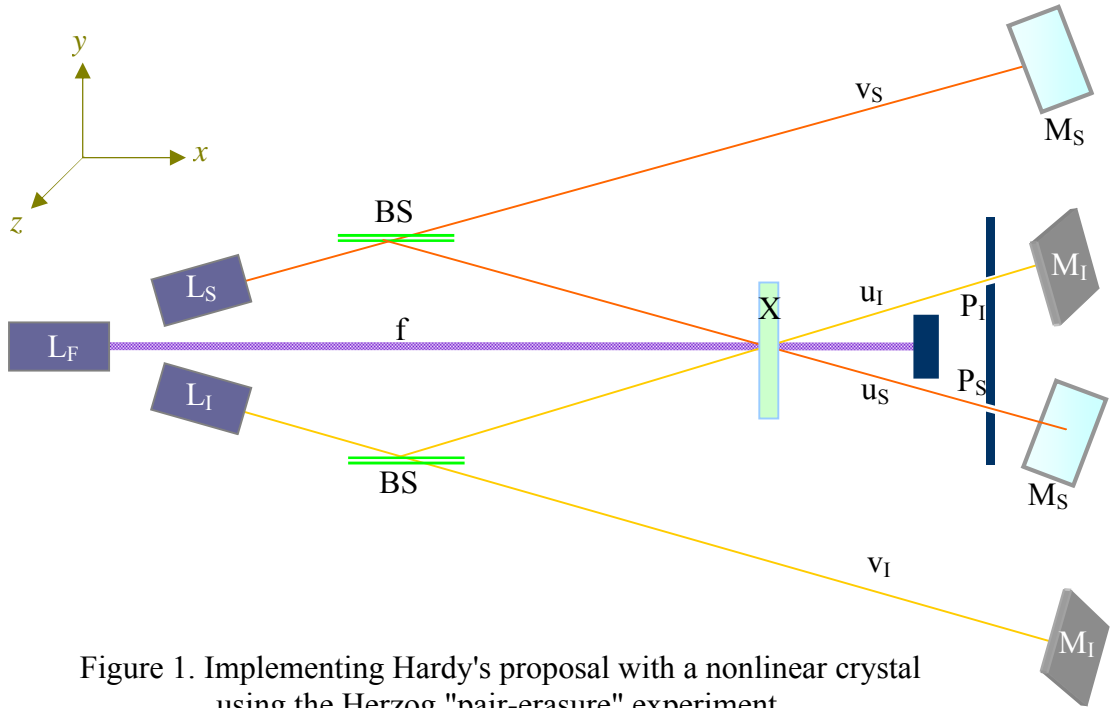

Figure 1. Implementing Hardy's proposal with a nonlinear crystal using the Herzog "pair-erasure" experiment.

the phases and intensities of the three lasers

(5) $\alpha\beta = 2q\gamma$ .

It may be achieved by adjusting the intensity of $L_S$ and $L_I$ to a sufficiently low value – since $|q|$ is a very small quantity, see the Appendix, and by monitoring the lasers' phases and discarding the detections obtained under phases that disobey (5).
Introducing (5) in (4) and renormalizing, one gets the Hardy state

(6) $|\Phi\rangle = 3^{-\frac{1}{2}}(\iota|u_S\rangle|v_I\rangle + \iota|v_S\rangle|u_I\rangle + |v_S\rangle|v_I\rangle)$ .

Fig. 2 describes the subsequent evolution of the beams $u_S$, $u_I$, $v_S$, $v_I$. The mirrors $M_I$, see also fig. 1, reflect the idler beams perpendicularly to the page in the negative direction of the axis $z$ and the mirrors $M_S$ reflect the signal beams perpendicularly to the page in the positive direction of the axis $z$. The beam-splitters $BS_I$ and $BS_S$ induce the transformations

(7) $|u\rangle = 2^{-\frac{1}{2}}(|c\rangle + \iota|d\rangle)$ ,   $|v\rangle = 2^{-\frac{1}{2}}(\iota|c\rangle + |d\rangle)$ .

Introducing (7) in (6) one gets

(8) $|\Phi\rangle = 12^{-\frac{1}{2}}(-3|c_S\rangle|c_I\rangle + \iota|c_S\rangle|d_I\rangle + \iota|d_S\rangle|c_I\rangle - |d_S\rangle|d_I\rangle)$ ,

in which one can see the cases $|d_S\rangle|d_I\rangle$, discussed in the next section.

### 3. Discussion

The clash between Hardy's experiment and the concept of trajectory is obvious for the cases of simultaneous detection in $D_S$ and $D_I$. Introducing the transformation (7) in (6) only for the signal photon, one gets

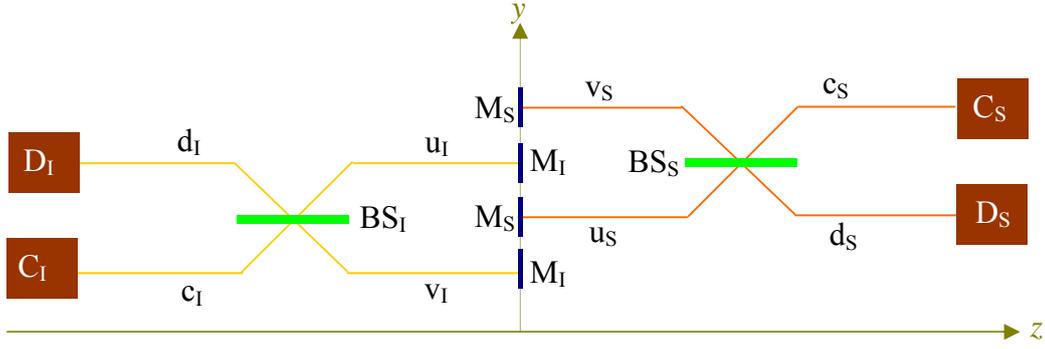

Figure 2. The motion in the plane $y$-$z$.

(9) $|H\rangle = 6^{-\frac{1}{2}}(2\iota|c_S\rangle|v_I\rangle - |c_S\rangle|u_I\rangle + \iota|d_S\rangle|u_I\rangle)$ .

One can conclude that from the point of view of an observer in movement with a suitable velocity in the direction $+z$, by the time the detector $D_S$ clicks the idler photon could be only on the track $u_I$. However, introducing in (6) the transformation (7) for the idler photon only, one gets

(10) $|H\rangle = 6^{-\frac{1}{2}}(2\iota|v_S\rangle|c_I\rangle - |u_S\rangle|c_I\rangle + \iota|u_S\rangle|d_I\rangle)$ ,

implying that for an observer in movement with the same velocity in the direction $-z$, by the time the detector $D_I$ clicks, the signal photon could be only on the track $u_S$.
Considering quicker and quicker traveling observers, there results that the idler photon should have traveled along $u_I$ and the signal photon along $u_S$. But there is no coupling $|u_S\rangle|u_I\rangle$ in (6).

We got *a clash* with the conclusion of the experiment described in introduction. Hardy's experiment seems to contradict the concept of trajectory, while the experiment in the introduction seems to leave no other choice than admitting trajectories. As to the possibility that during the measurements the particles entertain some "hidden communication" forward and backward in time, and agree to which tracks to settle, with particles that never met this is a problematic idea.

## Appendix

The down-conversion transformation is $\hat{U} = \exp(\iota\hat{H}\tau/\hbar)$, with the interaction Hamiltonian

(11) $\hat{H} = -\iota\hbar(\eta\,\hat{a}_F\,\hat{a}_S^\dagger\,\hat{a}_I^\dagger - \eta^*\,\hat{a}_F^\dagger\,\hat{a}_S\,\hat{a}_I)$ .

$\hat{U}$ may be expanded in the form

(12) $\hat{U} = 1 + q\,\hat{a}_F\,\hat{a}_S^\dagger\,\hat{a}_I^\dagger - q^*\,\hat{a}_F^\dagger\,\hat{a}_S\,\hat{a}_I + O(q^2)$ ,  $q = \eta\tau$ ,

where $q$ is the down-conversion amplitude. Since $|q|$ is a very small value (of order $10^{-6}$), the term in $q^2$ may be neglected. Moreover, this term is irrelevant since we study here only single signal-idler pairs.
Introducing (2) in (1) and applying (12) one gets

(13) $|\psi\rangle = N \{[|0_S\rangle + 2^{-\frac{1}{2}}\alpha|0_I\rangle(|v_S\rangle + \iota|u_S\rangle)][|0_I\rangle + 2^{-\frac{1}{2}}\beta|0_S\rangle(|v_I\rangle + \iota|u_I\rangle)](|0_F\rangle + q\gamma|1_S\rangle|1_I\rangle)$

$\times (\sum_0^\infty \gamma^n (n!)^{-\frac{1}{2}}|n_F\rangle) + q^*(\alpha\beta/2) |0_S\rangle|0_I\rangle(|1_F\rangle + 2^{-\frac{1}{2}}\gamma|2_F\rangle + 6^{-\frac{1}{2}}\gamma^2 |3_F\rangle + \ldots)\}$.

Due to the selection of down-conversion photons of the same wavelengths as those from $L_S$ and $L_I$, see fig. 1, one may replace in the last product on the first line $|1_S\rangle$ by $|u_S\rangle$, and $|1_I\rangle$ by $|u_I\rangle$.
Opening the parentheses,

(14) $|\psi\rangle = N \{|0_S\rangle|0_I\rangle|0_F\rangle + q^*(\alpha\beta/2) |0_S\rangle|0_I\rangle(|1_F\rangle + 2^{-\frac{1}{2}}\gamma|2_F\rangle + 6^{-\frac{1}{2}}\gamma^2 |3_F\rangle + \ldots)$

$+ 2^{-\frac{1}{2}}[\alpha|0_I\rangle(|v_S\rangle + \iota|u_S\rangle) + \beta|0_S\rangle(|v_I\rangle + \iota|u_I\rangle)]|0_F\rangle$

$+ (\alpha\beta/2)[(|v_S\rangle + \iota|u_S\rangle)(|v_I\rangle + \iota|u_I\rangle)|0_F\rangle + (2q\gamma/\alpha\beta)|0_S\rangle|0_I\rangle|u_S\rangle|u_I\rangle]$

$+ 2^{-\frac{1}{2}} q\gamma\alpha[|0_I\rangle(|v_S\rangle + \iota|u_S\rangle) + (\beta/\alpha)0_S\rangle(|v_I\rangle + \iota|u_I\rangle)]|u_S\rangle|u_I\rangle$

$+ (q\gamma\alpha\beta/2) (|v_S\rangle + \iota|u_S\rangle)(|v_I\rangle + \iota|u_I\rangle)|u_S\rangle|u_I\rangle\}(\sum_0^\infty \gamma^n (n!)^{-\frac{1}{2}}|n_F\rangle)$

The first two lines contain no signal-idler pairs so they are of no relevance for our study. On the third there appear single such pairs, on the fourth line three low-energy photons, and on the fifth line two pairs together. According to the constraints expressed in section 2, $|\alpha| = |\beta|$ and $\alpha\beta = 2q\gamma$, (condition (5) ), one can see that the amplitude of probability of three low-energy photons is one order of magnitude smaller than for single pairs, and the amplitude of probability for two pairs is two orders of magnitude smaller than for single pairs. So the lines fourth and fifth represent negligible noise, and we may retain for the description of the pair-production the expression on the third line

(15) $(|v_S\rangle + \iota|u_S\rangle)(|v_I\rangle + \iota|u_I\rangle)|0_F\rangle + (2q\gamma/\alpha\beta)|0_S\rangle|0_I\rangle|u_S\rangle|u_I\rangle$.